\documentclass[preprint,12pt,authoryear]{elsarticle}

\usepackage{amssymb}
\usepackage{lipsum}

\usepackage{adjustbox}
\usepackage{amsmath}
\usepackage{url}

\def\azh{\ref@jnl{AZh}}                 
\def\baas{\ref@jnl{BAAS}}               
\def\bac{\ref@jnl{Bull. astr. Inst. Czechosl.}}

\def\jcap{\ref@jnl{J. Cosmology Astropart. Phys.}}
\def\jrasc{\ref@jnl{JRASC}}             
\def\memras{\ref@jnl{MmRAS}}            
\def\nar{\ref@jnl{New A Rev.}}          
\def\pra{\ref@jnl{Phys.~Rev.~A}}        
\def\prb{\ref@jnl{Phys.~Rev.~B}}        
\def\prc{\ref@jnl{Phys.~Rev.~C}}        
\def\prd{\ref@jnl{Phys.~Rev.~D}}        
\def\pre{\ref@jnl{Phys.~Rev.~E}}        
\def\prl{\ref@jnl{Phys.~Rev.~Lett.}}    
\def\rmxaa{\ref@jnl{Rev. Mexicana Astron. Astrofis.}}%
\def\qjras{\ref@jnl{QJRAS}}             
\def\skytel{\ref@jnl{S\&T}}             
\def\solphys{\ref@jnl{Sol.~Phys.}}      
\def\sovast{\ref@jnl{Soviet~Ast.}}      
\def\ssr{\ref@jnl{Space~Sci.~Rev.}}     
\def\zap{\ref@jnl{ZAp}}                 
\def\iaucirc{\ref@jnl{IAU~Circ.}}       

\def\bain{\ref@jnl{Bull.~Astron.~Inst.~Netherlands}} 
\def\fcp{\ref@jnl{Fund.~Cosmic~Phys.}}  
\def\gca{\ref@jnl{Geochim.~Cosmochim.~Acta}}   
\def\grl{\ref@jnl{Geophys.~Res.~Lett.}} 
\def\jcp{\ref@jnl{J.~Chem.~Phys.}}      
\def\jgr{\ref@jnl{J.~Geophys.~Res.}}    
\def\jqsrt{\ref@jnl{J.~Quant.~Spec.~Radiat.~Transf.}}
\def\memsai{\ref@jnl{Mem.~Soc.~Astron.~Italiana}}
\def\nphysa{\ref@jnl{Nucl.~Phys.~A}}   
\def\physrep{\ref@jnl{Phys.~Rep.}}   
\def\physscr{\ref@jnl{Phys.~Scr}}   

\newcommand{\beq}{\begin{equation}}
\newcommand{\eeq}{\end{equation}}
\newcommand{\beqn}{\begin{eqnarray}}
\newcommand{\eeqn}{\end{eqnarray}}

\journal{New Astronomy}

\begin{document}

\begin{frontmatter}


\title{Transit Timing Variations of the Sub-Saturn Exoplanet HAT-P-12b}

\author[label1]{Kaviya Parthasarathy}
\ead{kaviyasarathy1998@gmail.com}
\author[label1]{Hsin-Min Liu}
\ead{shayna501@gapp.nthu.edu.tw}
\author[label1,label2]{Ing-Guey Jiang}
\ead{jiang@phys.nthu.edu.tw}
\author[label3]{Li-Chin Yeh}
\ead{lichinyeh@mx.nthu.edu.tw}
\author[label4]{Napaporn A-thano}
\author[label4]{Supachai Awiphan}
\author[label1]{Wen-Chi Cheng}
\author[label1]{Devesh P. Sariya}
\author[label5]{Shraddha Biswas}
\author[label5]{Devendra Bisht}
\author[label6]{Evgeny Griv}
\author[label4]{David Mkrtichian}
\author[label7]{Vineet Kumar Mannaday}
\author[label8]{Parijat Thakur}
\author[label9]{Aleksey Shlyapnikov}

\address[label1]{Institute of Astronomy, National Tsing Hua University, Hsin-Chu, Taiwan} 
\address[label2]{Department of Physics, National Tsing Hua University, Hsin-Chu, Taiwan} 
\address[label3]{Institute of Computational and Modeling Science, National Tsing Hua University, Hsin-Chu, Taiwan}
\address[label4]{National Astronomical Research Institute of Thailand, Chiang Mai, 50180, Thailand}
\address[label5]{Indian Centre for Space Physics, 466 Barakhola, Singabari Road, Netai Nagar, Kolkata, West Bengal 700099, India
}
\address[label6]{Department of Physics, Ben-Gurion University, Beer-Sheva 84105, Israel}
\address[label7]{
Department of Physics, Govt. Niranjan Kesharwani College, Kota, Bilaspur (C.G.)-495113, India
}
\address[label8]{Department of Pure \& Applied Physics, Guru Ghasidas Vishwavidyalaya (A Central University), Bilaspur (C.G.) - 495009, India}
\address[label9]{Crimean Astrophysical Observatory, Russian Academy of Sciences, 
298409,
Nauchny, Crimea, Russia}

\begin{abstract}
We present Transit Timing Variations (TTVs) of HAT-P-12b,
a low-density sub-Saturn mass planet orbiting a metal-poor
K4 dwarf star. 
Using 14 years of observational data (2009-2022), our study incorporates 7 new ground-based photometric transit observations, three sectors of Transiting Exoplanet Survey Satellite (TESS) data, and 23 previously published light curves.
A total of  46 light curves were analyzed using various analytical models, such as linear, orbital decay, apsidal precession, and sinusoidal models to investigate the presence of additional planets. 
The stellar tidal quality factor 
($Q_\star' \sim$ 28.4) is lower than the theoretical predictions, making the orbital decay model an unlikely explanation. The apsidal precession model with a $\chi_r^2$ of 4.2 revealed a slight orbital eccentricity 
(e = 0.0013) and a precession rate of 0.0045 rad/epoch. Frequency analysis using the Generalized Lomb-Scargle (GLS) periodogram identified a significant periodic signal at 0.00415 cycles/day 
(FAP = 5.1$\times$10$^{-6}$ \%), 
suggesting the influence of an additional planetary companion. The sinusoidal model provides the lowest reduced chi-squared value ($\chi_r^2$) of 3.2. Sinusoidal fitting of the timing residuals estimated this companion to have a mass of approximately 0.02 $M_J$ , assuming it is in a 2:1 Mean-Motion Resonance
(MMR) with HAT-P-12b. Additionally, the Applegate mechanism, with an amplitude much smaller than the observed TTV amplitude of 156 s, confirms that stellar activity is not responsible for the observed variations.

\end{abstract}


\end{frontmatter}

\section{Introduction}
Exoplanet research has transformed our understanding of planets beyond our solar system and has enhanced the quest for extraterrestrial life by identifying potentially habitable planets.  
As noted in the NASA Exoplanet Archive\footnote{\url{https://exoplanetarchive.ipac.caltech.edu/}}, approximately 6000 exoplanets had been identified through both ground-based and space-based telescopes by January 2025.  
Most of these discoveries have been made possible through observations of planetary transits and radial velocity (RV). Various dedicated space missions, including Kepler, K2, and TESS, and ground-based programs, including HATNet, SuperWASP, XO, and others, have contributed to the discovery
\citep{2004PASP..116..266B,2005PASP..117..783M, 2006PASP..118.1407P,2011ApJ...728..117B}.  
A planetary transit occurs when a planet crosses in front of its host star, with its orbit nearly aligned with our line of sight.  
The resulting dip in the light curve allows us to determine the ratio of the planet's radius to that of the host star (\textit{$R_p/R_\star$}), the semi-major axis (\textit{a}), the orbital inclination (\textit{i}), and the orbital period (\textit{P}) from the interval between two consecutive transits \citep{2002ApJ...580L.171M,2013PASP..125...83E}. RV measurements of the host star allow us to ascertain the planet's mass as well as the eccentricity of its orbit \citep{2010ARA&A..48..631S}.
Therefore, spectroscopic and photometric observations are essential for understanding the physical properties of the planetary system. 
Transit Timing Variations (TTVs) are an effective tool for detecting and characterizing exoplanetary systems. By examining minute deviations in the timing of a planet's transit, we can deduce the existence of other planets or explore intricate orbital dynamics \citep{2005MNRAS.359..567A,2005Sci...307.1288H,2010MNRAS.407.2625M,2013AJ....145...68J}. These signals are particularly prominent in systems where planets are in or near Mean-Motion Resonances (MMR), facilitating precise measurements of the mass and orbital characteristics of the influencing planet.
TTVs provides insights into orbital changes, such as orbital decay; occurs due to tidal dissipation between a planet and its host star, leading to a gradual decrease in the orbital period \citep{2016AJ....151...17J,2024NewA..10602130Y} and Apsidal Precession; changes in the orientation of a planet’s orbit over time due to gravitational interactions \citep{1995Ap&SS.226...99G}. Several authors investigated the theoretical predictions of TTVs \citep{2016A&A...588L...6M,2017AJ....154....4P,2019MNRAS.490.4230S,2020AJ....160...47M,2020AAS...23545602Y, 2022AJ....163...77A, Mannaday2022}.

In 2006, HAT-P-12b was discovered using the HAT-5 telescope as part of the Hungarian Automated Telescope Network \citep{2004PASP..116..266B}. Subsequent photometric observations by \cite{2009ApJ...706..785H} indicated that HAT-P-12b is a low-density warm sub-Saturn with a mass of $ M_p = 0.211 M_J $, a radius of $ R_p = 0.959 R_J $, and a density of $\rho_p = 0.0295 $ g cm$^{-3}$, with an orbital period of 3.21 days. It transits its host star, HAT-P-12, which is a relatively metal-poor K4 dwarf star with [Fe/H] $= -0.29$. The orbital and physical parameters of HAT-P-12 system have been refined by several studies (see for e.g., \cite{2015A&A...583A.138M, 2012AJ....143...95L, 2018A&A...620A.142A, 2018A&A...613A..41M}). The stellar and planetary parameters used in our studies are provided in Table \ref{table:fundamental_values}. There have been no observed occultations of HAT-P-12b by its star \citep{2013ApJ...770..102T}. Several authors investigated the atmospheric characteristics of HAT-P-12b. However, photometric examinations of HAT-P-12b have been confined to a limited studies, such as \cite{2019MNRAS.486.2290O, 2021RAA....21...97S, 2023AcA....73..159M}. The studies of \cite{2019MNRAS.486.2290O, 2023AcA....73..159M} discovered no evidence of a transiting planetary companion. However, \cite{2021RAA....21...97S} found non-sinisoidal TTVs in HAT-P-12b. Their two-planet model indicates the presence of a 0.2 $M_J$ companion in an 8.8-day orbit. The contradicting evidence emphasizes the necessity for additional inquiry. A long-term, high-precision monitoring effort is required to increase the accuracy of transit data.

In this paper, we introduce 7 newly observed ground-based transit lightcurves of HAT-P-12b collected across different epoch. These new observations are combined with, three sectors of Transiting Exoplanet Survey Satellite (TESS) and 23 publicly available lightcurves, spanning a total baseline of 14 years. We aim to determine whether the observed Transit Timing Variations(TTVs) of HAT-P-12b are caused by a planetary companion or by other factors. 

Section \ref{sec:data} provides details on observations, data reduction techniques, and an overview of TESS data and light curves from the literature. Section \ref{sec:lightcurve} discusses the analysis of light curves to determine orbital parameters. The section \ref{sec:TTV} provides a detailed examination of TTVs using various models. The summary and concluding remark are provided in section \ref{sec:RD}.

\begin{table}[ht]
\caption{The observational record of HAT-P-12b.}
\vspace{0.2cm}
\label{table:Observation}
\begin{adjustbox}{width=1.0\textwidth}
\begin{tabular}{ccccccc}
\hline
\hline
Date & Epoch & Telescope & Filter & Duration of Exposure(s) & No. of Images & Transit Coverage \\ \hline
\hline
2012/06/01 & 580 & WISE & R & 20 & 218 & Full \\ 
2014/04/17 & 793 & P60 & R & 20 & 219 & Full \\ 
2015/05/01 & 911 & P60 & R & 30 & 146 & Egress only \\ 
2015/06/15 & 925 & P60 & R & 24 & 96 & Egress only \\ 
2022/01/09 & 1672 & TRT & R & 30 & 331 & Full \\ 
2022/02/07 & 1681 & TRT & R & 30 & 221 & Ingress only \\ 
2022/05/08 & 1709 & TRT & R & 30 & 400 & Full \\ 
\hline
\end{tabular}
\end{adjustbox}
\end{table}

\section{Data \label{sec:data}}

\subsection{Ground-based observations}

We conducted 7 new photometric observations of HAT-P-12b using three different telescopes: \\

    1. The 60-inch telescope (P60) at Palomar Observatory, USA. \\
    
    2. The 0.7m Thai Robotic Telescope at Sierra Remote Observations
    (TRT-SRO) in California, USA, equipped with an Andor iKon-M 934 CCD camera.\\
    
    3. The 1m telescope at Wide Field Infrared Survey Explorer(WISE) Observatory, Israel. 
    
The details of these observations, including the date, epoch, telescope used, filter type, and exposure times are summarized in Table \ref{table:Observation}.

\subsection{Data reductions}
\begin{figure*}[htbp]
    \centering
    \includegraphics[width=0.45\textwidth,
    height=3.7cm]{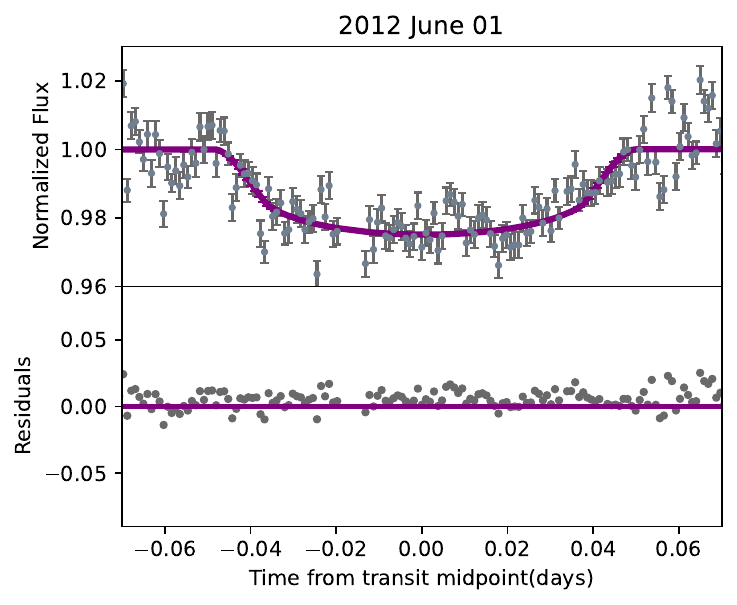}
    \hspace{0.05\textwidth} 
    \includegraphics[width=0.45\textwidth,
    height=3.7cm]{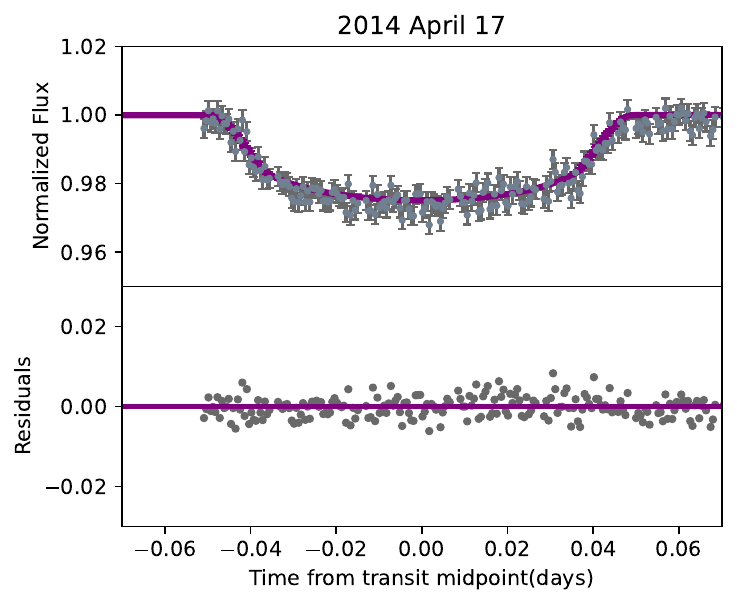}

    \includegraphics[width=0.45\textwidth, height=3.7cm]{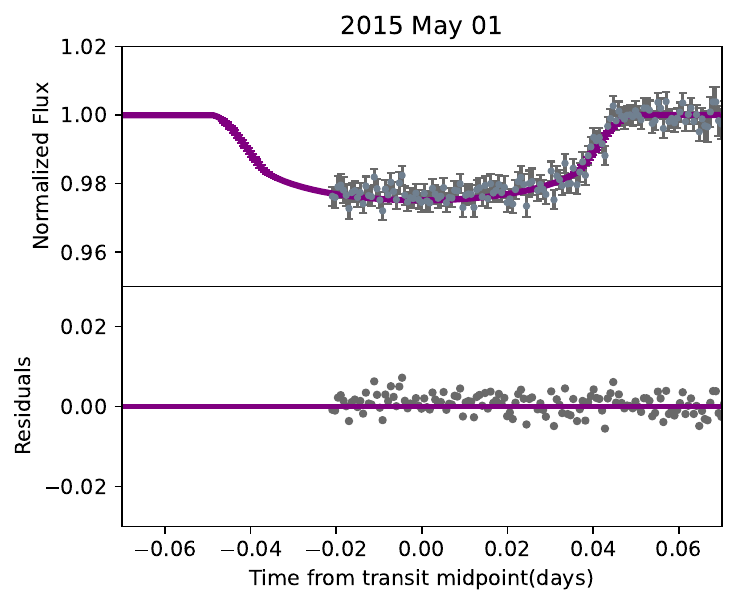}
    \hspace{0.05\textwidth}
    \includegraphics[width=0.45\textwidth, height=3.7cm]{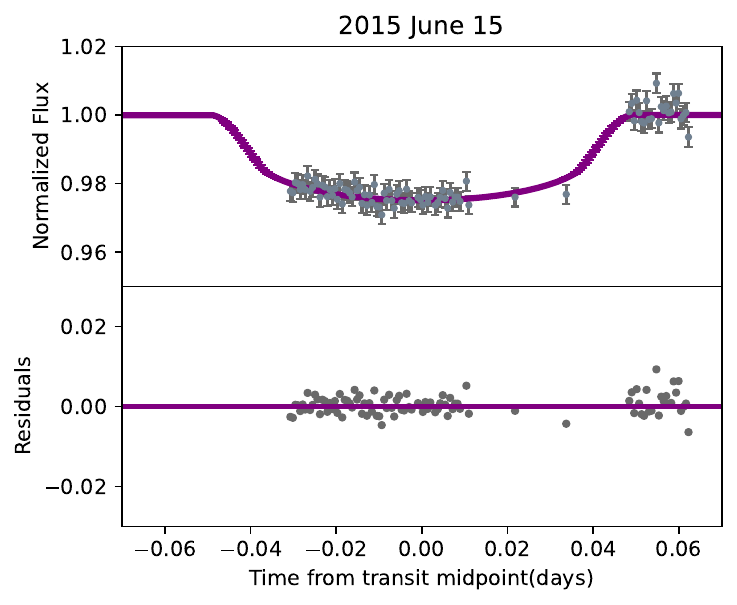}


    \includegraphics[width=0.45\textwidth, height=3.7cm]{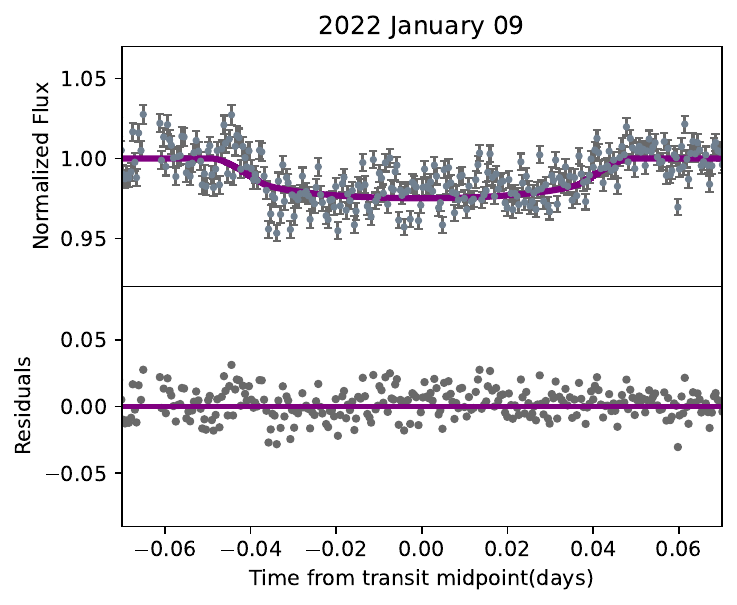}
    \hspace{0.05\textwidth}
    \includegraphics[width=0.45\textwidth, height=3.7cm]{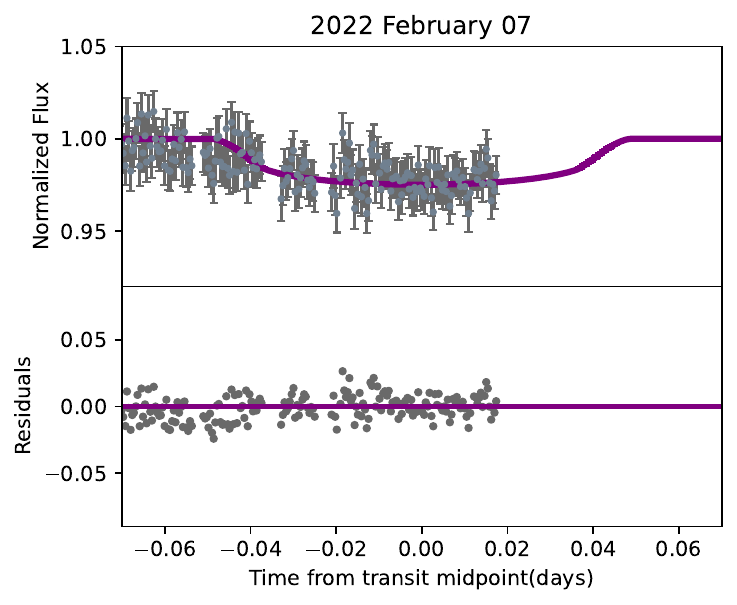}


    \includegraphics[width=0.45\textwidth, height=3.7cm]{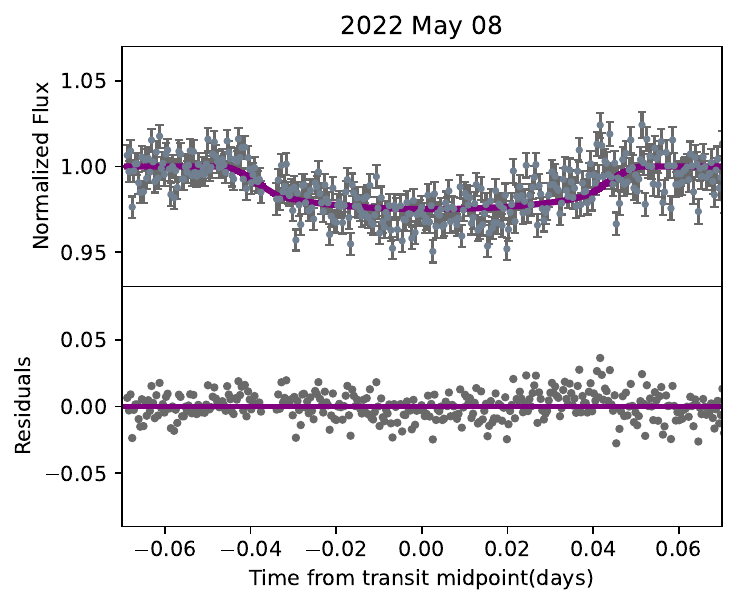}
    
    \caption{Ground-based transit light curves of HAT-P-12b from various observation epochs. The solid lines represent the best-fit models, and the residuals are shown below each light curve.}
    \label{fig:ourstudy}
\end{figure*}

The raw CCD images from the three telescopes were processed using 
IRAF\footnote{IRAF is distributed by the National Optical Astronomy Observatories, which are operated by the Association of Universities for Research in Astronomy, Inc., under a cooperative agreement with the National Science Foundation.\url{http://iraf.noao.edu/}}software for data reduction. The reduction process begins with bias correction, dark correction, and flat-field correction and calibrated for photometric analysis.
Aperture photometry was performed using the IRAF \texttt{phot} package. Aperture sizes were optimized based on the seeing conditions to maximize the signal-to-noise ratio. Two or three comparison stars, with a magnitude similar to HAT-P-12 were visually inspected and chosen. The \texttt{txt$_{dump}$} task was used to extract the results of aperture photometry which are the times, fluxes, and magnitude measurements of the target star and the comparison stars. The $ BJD_{TDB} $, relative flux, and its errors were used to generate light curves. Relative flux and its associated errors were calculated based on the flux and magnitude values of the target and the comparison stars provided by IRAF. 
The observation times were converted to Barycentric Julian Date in Barycentric Dynamical Time ($BJD_{TDB}$) using an online tool \footnote{
 \url{https://astroutils.astronomy.osu.edu/time/utc2bjd.html}} developed by \cite{2010PASP..122..935E}.
 The normalized light curves and their residuals obtained from our observations are illustrated in Figure \ref{fig:ourstudy}. The normalized light curves can be accessed through a machine-readable format shown in Table \ref{table:tessmachine}.

\subsection{Transiting exoplanet survey satellite (TESS)}
\begin{figure*}
    \centering
    \includegraphics[width=0.23\textwidth,
    height=3.5cm]{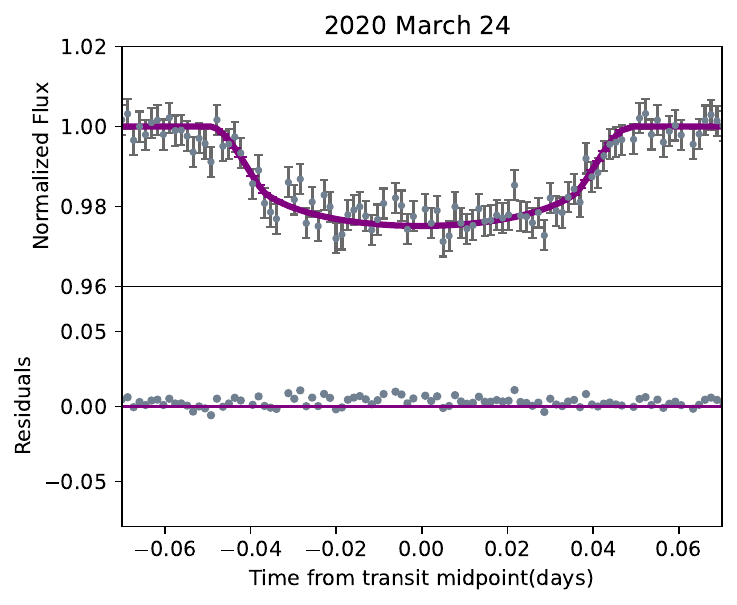}
    \hspace{0.0\textwidth} 
    \includegraphics[width=0.23\textwidth,
    height=3.5cm]{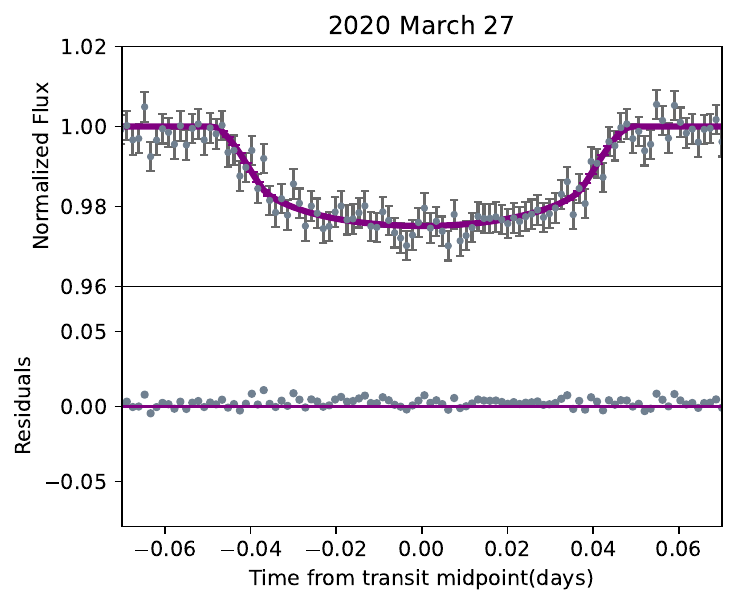}
    \hspace{0.0\textwidth}
    \includegraphics[width=0.23\textwidth,
    height=3.5cm]{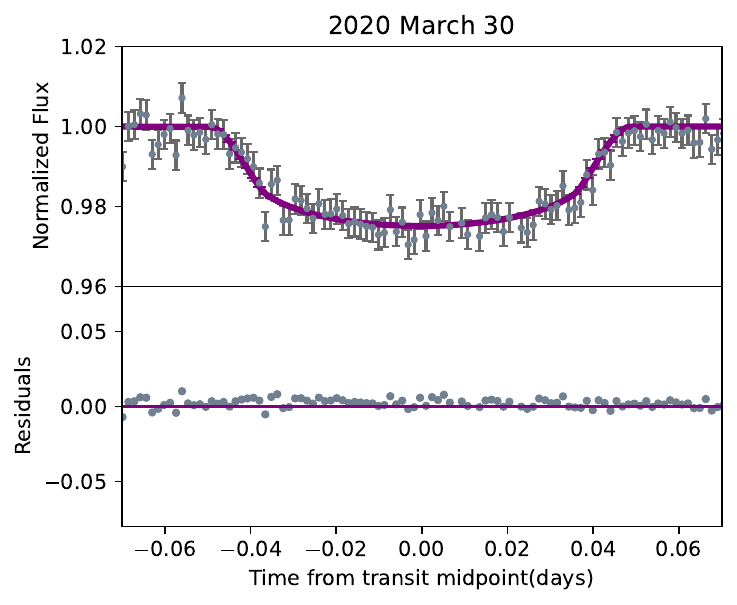}
    \hspace{0.0\textwidth}
    \includegraphics[width=0.23\textwidth,
    height=3.5cm]{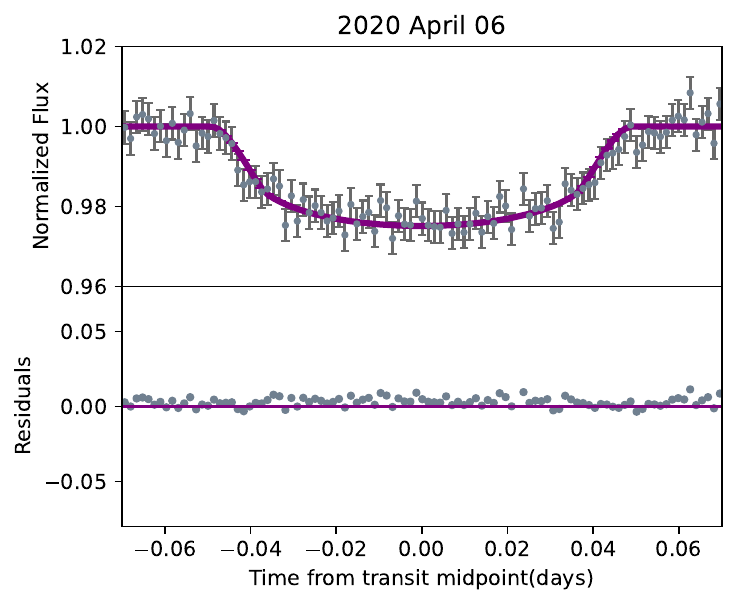}

    \vspace{0.5cm} 

    \includegraphics[width=0.23\textwidth, height=3.5cm]{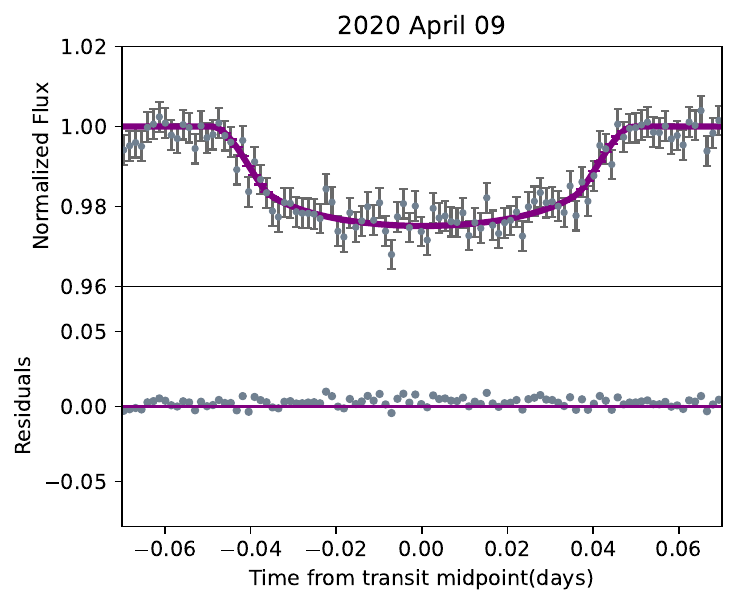}
    \hspace{0.0\textwidth}
    \includegraphics[width=0.23\textwidth, height=3.5cm]{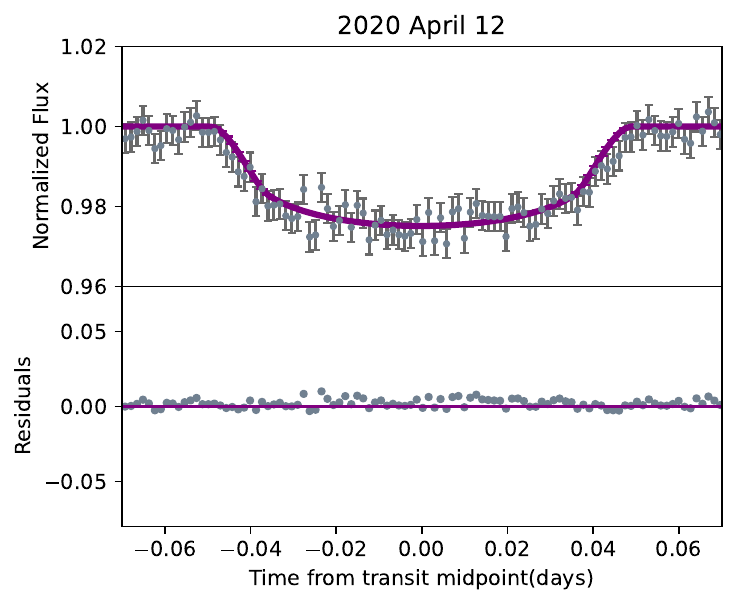}
    \hspace{0.0\textwidth}
    \includegraphics[width=0.23\textwidth, height=3.5cm]{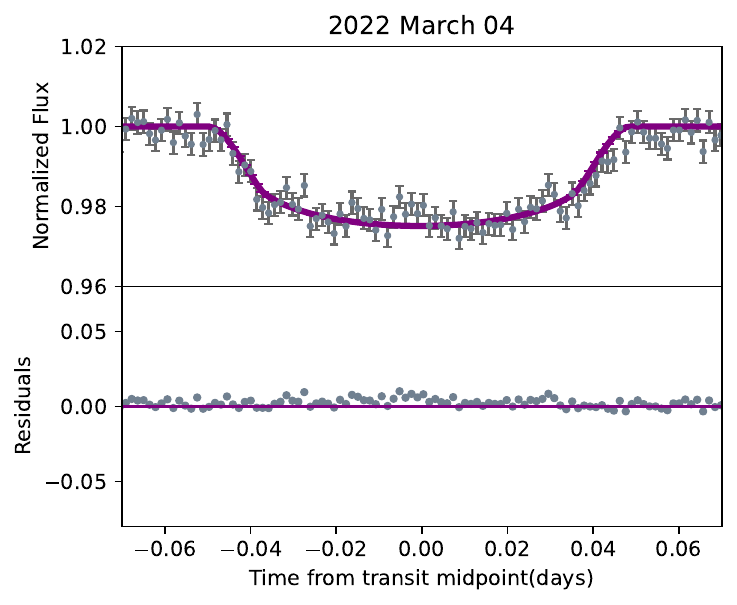}
    \hspace{0.0\textwidth}
    \includegraphics[width=0.23\textwidth, height=3.5cm]{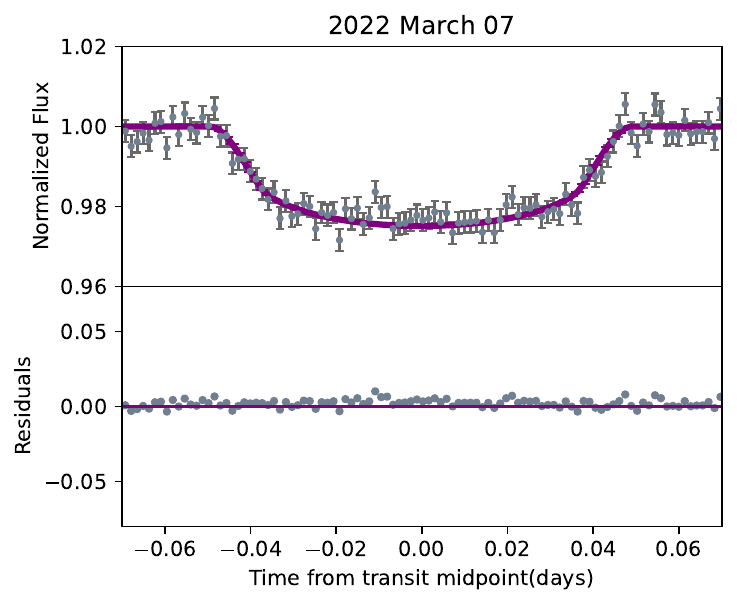}

    \vspace{0.5cm}
    
    \includegraphics[width=0.23\textwidth, height=3.5cm]{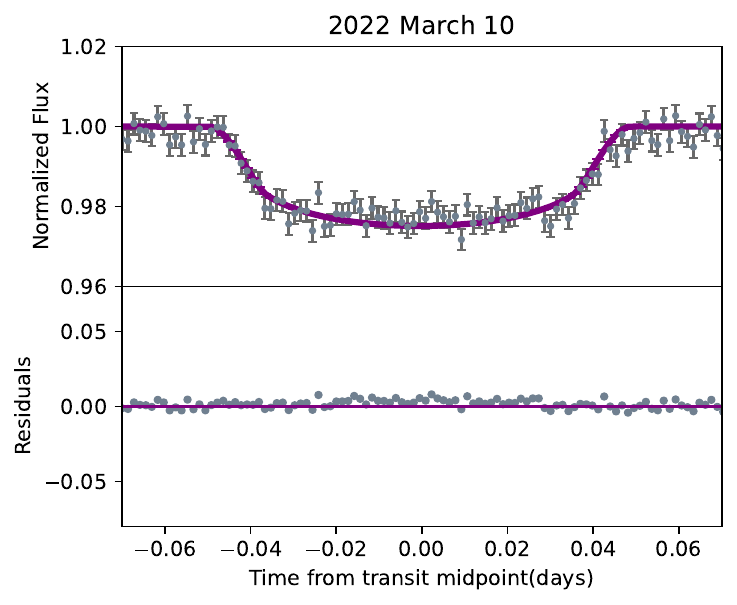}
     \hspace{0.0\textwidth}
    \includegraphics[width=0.23\textwidth, height=3.5cm]{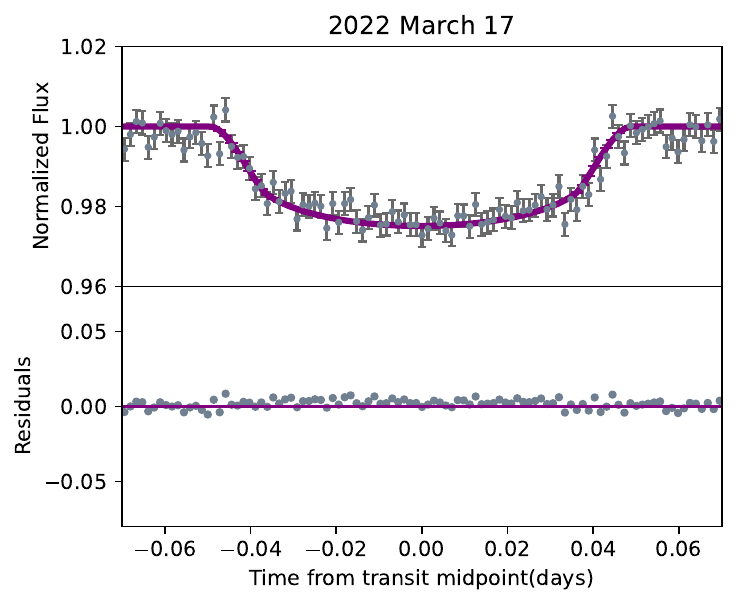}
    \hspace{0.0\textwidth}
    \includegraphics[width=0.23\textwidth, height=3.5cm]{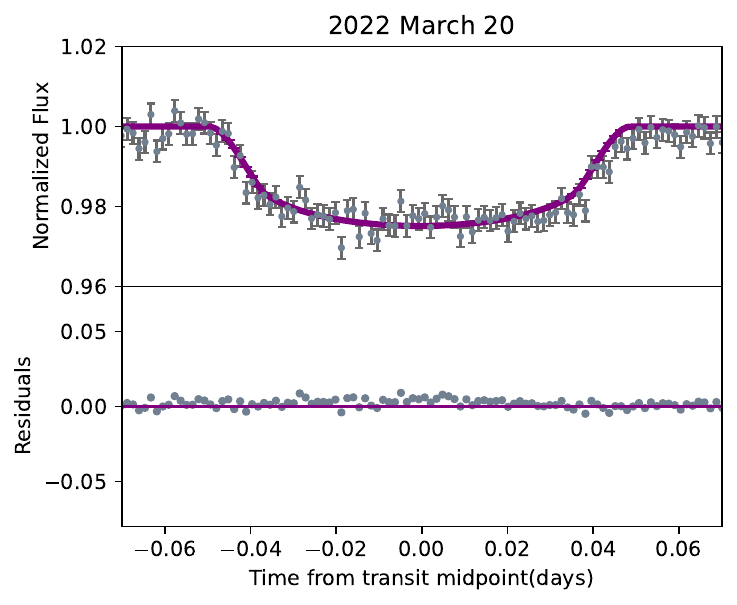}
     \hspace{0.0\textwidth}
    \includegraphics[width=0.23\textwidth, height=3.5cm]{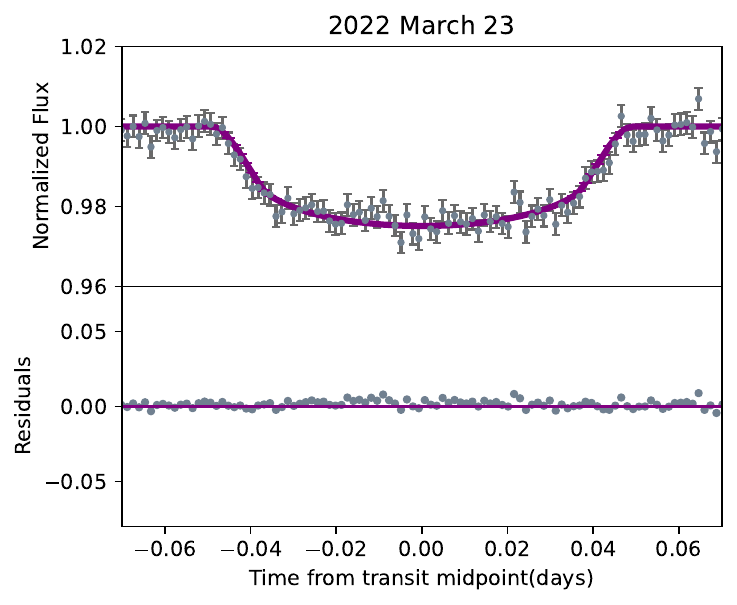}
    \vspace{0.5cm}
    
    \includegraphics[width=0.23\textwidth, height=3.5cm]{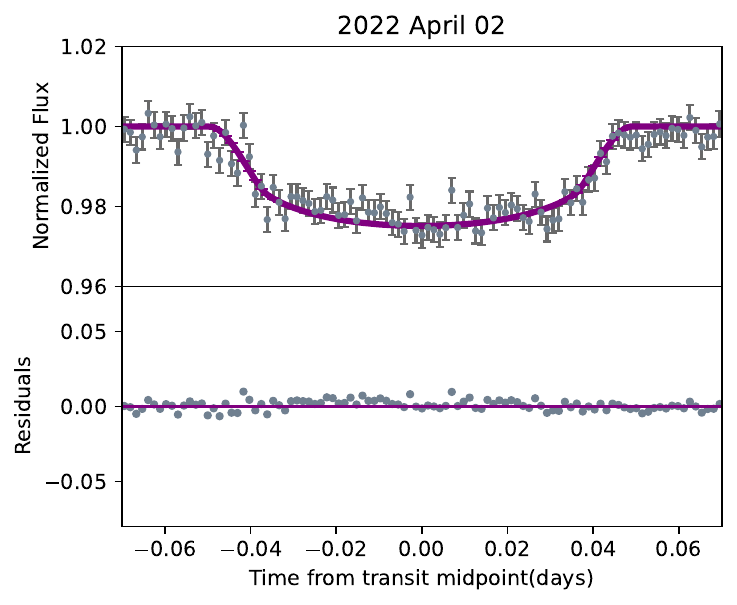}
    \hspace{0.0\textwidth} 
    \includegraphics[width=0.23\textwidth, height=3.5cm]{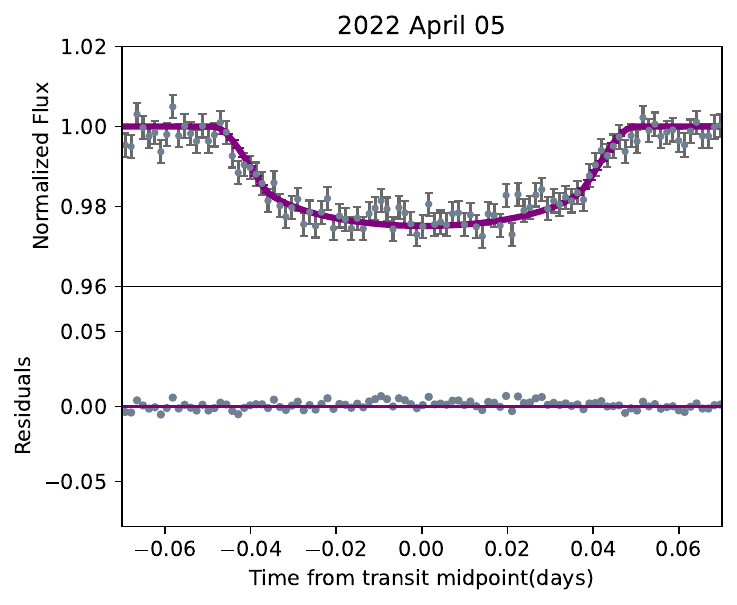}
    \hspace{0.0\textwidth}
    \includegraphics[width=0.23\textwidth, height=3.5cm]{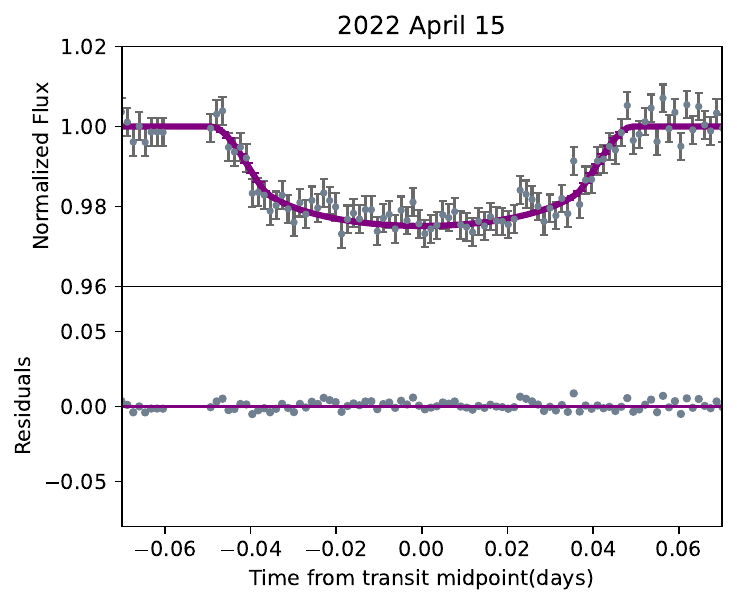}
    \hspace{0.0\textwidth}
    \includegraphics[width=0.23\textwidth, height=3.5cm]{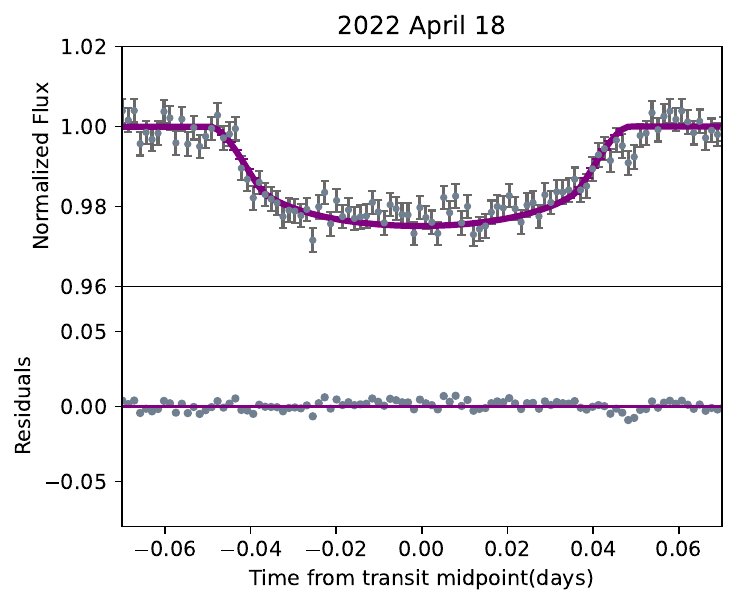}

    \caption{Space-based transit light curves of HAT-P-12b from different sectors of TESS. The solid lines represent the best-fit models, and the residuals are shown below each light curve.}
    \label{fig:TESS}
\end{figure*}

TESS observed HAT-P-12 in sectors 23,49 and 50 covering 16 transits.
We retrieved the Photometric data of these sectors, which are the times, fluxes, and errors from the  Mikulski
Archive for Space Telescope (MAST)
\footnote{ \url{http://mast.stsci.edu}}
using JULIET package \citep{2019MNRAS.490.2262E}. Sector 23 contains 6 transits covering from 2020/03/24 to 2020/04/12. Sector 49 contains 5 transits covering from 2022/03/04 to 2022/03/23. Sector 50 contains 5 transits covering from 2022/04/02 to 2022/04/18. 
Each transit observed by TESS provided over 3,000 data points. To determine the mid-transit time, we only require the transit portion of the light curve, therefore the out-of-transit baseline which primarily contains noise and stellar Variability was manually removed. The processed light curves for all 16 transits observed by TESS are presented in 
Figure \ref{fig:TESS} and the values are available in the machine-readable format shown in Table \ref{table:tessmachine}.

\begin{table}
\centering
\caption{The normalized light curves of HAT-P-12b obtained by our ground-based observations and TESS. A small portion is shown here.
The entire table is available in the machine-readable form.
}\label{table:tessmachine}
\vspace{0.3cm}
\begin{tabular}{cccc}
\hline
\hline
Epoch & BJD\textbf{$_{TDB}$} & Normalized Flux &  Uncertainty\\ \hline
\hline
580 &2456080.241507 & 1.00751 & 0.00394\\
    &2456080.242333 & 1.00402 & 0.00392\\
    &2456080.243159 & 1.01152 & 0.00395\\
    &2456080.243983 & 1.00774 & 0.00394\\
    &2456080.244808 & 1.00641 & 0.00393\\
    & - - - & - - - & - - -\\
793 & 2456764.676953 & 1.01611 &0.00281\\
    &2456764.677480 &1.01836 &0.00281\\
    &2456764.678005 &1.02127 &0.00282\\
    &2456764.678529 &1.01788& 0.00281\\
    &2456764.679055 &1.01889 & 0.00282\\
    
    & - - - & - - - & - - -\\    
911 & 2457143.848503& 0.99631 &0.00275\\
    & 2457143.849147& 0.99596 &0.00318\\
    & 2457143.849808& 0.99908 &0.00319\\
    & 2457143.850452& 0.99957 &0.00319\\
    & 2457143.851097& 0.99809 &0.00379\\
    & - - - & - - - & - - -\\
    \hline
\end{tabular}
\end{table}

\subsection{Literature data}
This work utilizes a total of 23 publicly available light curves (\cite{2009ApJ...706..785H},\cite{2012AJ....143...95L},\cite{2015A&A...583A.138M},\cite{2018A&A...620A.142A},\cite{2021RAA....21...97S}). The light curves are analyzed using the same method as our observation data (refer section \ref{sec:lightcurve}) to reconfirm their mid-transit times.

\section{Light Curve Analysis\label{sec:lightcurve}}

The orbital and planetary parameters of HAT-P-12b were derived using the Python package \texttt{pylightcurve}, specifically designed for modeling and analyzing light curves.
A total of 46 light curves were modeled individually. \texttt{pylightcurve} utilizes the  Exoplanet Characterisation Catalogue(ECC), developed as part of the Exoclock Project \citep{2022ExA....53..547K}. This ECC catalog contains information on more than 370 exoplanets including HAT-P-12b. Access to the data is facilitated by the \texttt{pylightcurve.get\_planet()} function, which provides all the necessary parameters and its initial corresponding values for modeling light curves. 
The adopted stellar and planetary parameters are listed in Table \ref{table:fundamental_values}.

\begin{table}[ht]
\centering
\caption{Stellar and planetary parameters of HAT-P-12b used in our work. The parameters are adopted from \citep{2009ApJ...706..785H}. } 
\label{table:fundamental_values}
\begin{tabular}{cc}
\hline
\hline

Parameter & Value  \\ \hline
\hline
\multicolumn{2}{c}{Stellar Parameters} \\ \hline
\hline
$M_{\star}(M_\odot)$ & $0.733 \pm 0.018$  \\ 
$R_{\star}(R_\odot)$ & $0.701^{+0.017}_{-0.012}$  \\ 
$\log{g}$ (cgs) & $4.61 \pm 0.01$  \\ 
$\text{[Fe/H]}$ & $-0.29 \pm 0.05$  \\ 
$T_{\text{eff}}$ (K) & $4650 \pm 60$  \\ 
$V$ (mag) & $12.84$  \\
\hline
\hline
\multicolumn{2}{c}{Planetary Parameters} \\ \hline
\hline
$M_{p}(M_J)$ & $0.211 \pm 0.012$  \\ 
$R_{p}(R_J)$ & $0.959^{+0.029}_{-0.021}$ \\ 
$a$ (AU) & $0.0384 \pm 0.0003$  \\ 
$\rho_p$ (g cm$^{-3}$) & $0.0295 \pm 0.025$ \\ \hline
\end{tabular}
\end{table}

The planetary parameters used by \texttt{pylightcurve} for modeling light curves include orbital period $ P $ in days, epoch of mid-transit $ T_0 $ in BJD, orbital inclination \textit{i} in degrees, eccentricity \textit{e}, semi-major axes \textit{a} in the units of stellar radius $ R_\star$, radius of the Planet $ R_p$ in the units of stellar radius $R_\star$, an argument of periastron $\omega$.
while fitting, the orbital period and eccentricity were treated as fixed parameters. The orbital period was fixed to enable the detection of TTVs, allowing for any observed deviations in transit timing.
The eccentricity was fixed, based on the assumption that the orbit of HAT-P-12b is circular. 
The remaining orbital parameters \textit{$t_0$, i, a} are set as free parameters. The parameter setting and the corresponding values used as inputs for modeling light curves are shown in Table \ref{table:Initial parameters}. 
 
The light curve fitting process was performed using the 
planet-transit-fitting function in \texttt{pylightcurve}, which employs an \textit{emcee} Markov Chain Monte Carlo (MCMC) \citep{2013PASP..125..306F} sampler to perform simultaneous detrending and normalization of the light curves. For detrending, a first-order polynomial function was applied to remove the linear trends in the baseline. 
The default configuration of this function, involves 130,000 steps, 200 walkers, and 30,000 burn-in steps. This gives adequate convergence of the chains and parameter exploration, providing posterior probability distributions. A quadratic  limb-darkening method was adopted to characterize the stellar limb-darkening effect. The filters used for observation and their associated limb darkening coefficients(\textit{$ u_1, u_2 $}), are shown in Table \ref{table:limbdarkening}. The overall fitting results are presented in Table \ref{table:midtime}.

\begin{table}[htbp]
\centering
\caption{The initial values and setting of planetary parameters used by the fitting processes of \texttt{pylightcurve}.}
\label{table:Initial parameters}
\begin{tabular}{cccc}
\hline
\hline
Parameters & Intial Values & Priors & Prior Distribution \\ \hline
\hline
P (days) & 3.21305762 & -- &Fixed \\
$ T_0 $(BJD$_{TDB}$) & 2456851.481119 & [-0.2, 0.2] & Uniform \\
$ i $ (deg) & 89.0 & [70.0, 90.0] & Uniform \\
$ a/R_\star $ & 11.77 & [0.5, 2.0] & Uniform \\
$ R_p/R_\star $ & 0.1406 & [0.5, 2.0] &Uniform \\
e & 0.0 & --&Fixed \\
$\omega$\textbf{(deg)} & 0.0 & --&Fixed \\
\hline
\end{tabular}
\end{table}

\begin{table}
\centering
\caption{The limb darkening coefficients of HAT-P-12b adopted 
from \texttt{pylightcurve}.}
\label{table:limbdarkening}
\begin{tabular}{ccc}
\hline
\hline
Filter& $ u_1 $&$ u_2 $\\ 
\hline
\hline
B & 0.93739 & -0.07324\\
R & 0.59585 & 0.11564\\
I & 0.47749 & 0.13526\\
g'& 0.87745 & -0.03303\\
r'& 0.62543 & 0.11007\\
z'& 0.41697 & 0.14313\\
TESS & 0.48727 & 0.13392\\
\hline
\end{tabular} 
\end{table}

\begin{figure*}[ht]
    \centering
    \includegraphics[width=0.9\linewidth, height=10cm]{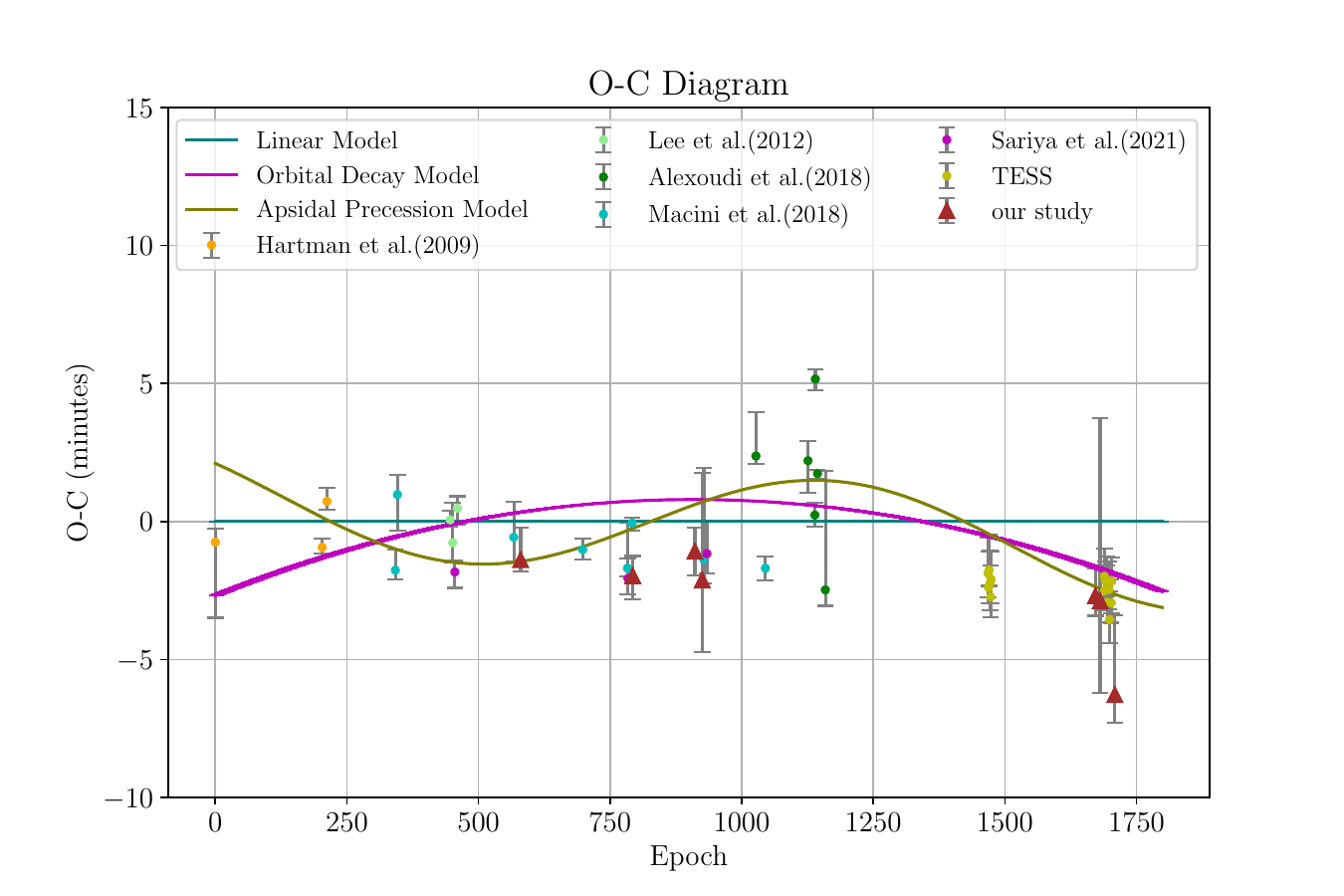}
    \caption{Transit Timing Variation (TTV) diagram for HAT-P-12b after subtracting the linear model. The plot compares the observed-minus-calculated (O-C) values against linear, orbital decay, and apsidal precession models}.
     \label{fig:ocplot}
\end{figure*}

\begin{table}
\centering
\caption{Fitting results obtained from \texttt{pylightcurve}}
\label{table:midtime}
\begin{adjustbox}{width=1.0\textwidth}
\begin{tabular}{cccccccc}
\hline
\hline
Epoch& $ Tm(BJD_{TDB}-2450000)$ (days) & Avg err  &  O-C(days)  & $ i\textbf{(\textdegree)}$ & $r_p/r_*$& $a/r_*$ & source \\ \hline
\hline
0&4216.77273 $_{-0.0019}^{+0.00035}$ &0.001125 & $-0.00053$ & 	88.9 $_{-0.19}^{+0.71}$ & 11.95 $ _{-0.20}^{+	0.08}$ & 0.1399 $ _{-0.0010}^{+0.0011}$ &	(a)\\
203	&4869.02361	$_{-0.00015	}^{+	0.00022} $ &	0.00019	&$	-0.00063$ & 89.1 $_{-1.0}^{+0.5}$ &	11.74 $ _{-0.68}^{+	0.14}$ &	0.143 $ _{-0.0017}^{+0.0026}$ &(a)\\
212	&4897.94230 $_{-0.00022	}^{+0.00033	} $ &	0.00028	&$	0.00053	$&	88.9	$_{-0.4	}^{+0.7	}$ &	12.39	$ _{-	0.34}^{+0.12}$ &	0.137	$ _{-0.0018	}^{+0.0030	}$ &	(a)\\
342	&5315.63826	$_{-0.00024	}^{+0.00053	} $ &	0.00039	&$	-0.00118$	&	89.1	$_{-0.3	}^{+0.7	}$ &	11.75	$ _{-	0.18}^{+0.37}$ &	0.138	$ _{-0.0023	}^{+0.0020	}$ &	(c)\\
346	&5328.4924		$_{-0.0009	}^{+0.0005	} $ &	0.0007	&$	0.00072	$ &	88.9	$_{-0.4	}^{+0.7	}$ &	11.93	$ _{-0.26}^{+0.13}$ &	0.1368	$ _{-0.0008	}^{+0.0021	}$ &	(c)\\
446	&5649.79768		$_{-0.00018	}^{+0.00024	} $ &	0.00021	&$	0.00010	$&	89.1	$_{-0.4	}^{+0.6	}$ &	11.68	$ _{-0.23}^{+0.09}$ &	0.1373	$ _{-0.0009	}^{+0.0018	}$ &	(d)\\
451	&5665.8624		$_{-0.0005}^{+0.0010} $ &	0.0008	&$	-0.00048	$&	89.0	$_{-0.6	}^{+0.6	}$ &	12.05	$ _{-0.48}^{+0.13}$ &	0.142	$ _{-0.0012	}^{+0.0029}$ &	(d)\\
455	&5678.7139		$_{-0.0004}^{+0.0003} $ &	0.0004	&$	-0.00121	$ &	89.1	$_{-0.3	}^{+0.7	}$ &	11.91	$ _{-0.22}^{+0.12}$ &	0.135	$ _{-0.0011	}^{+0.0021	}$ &	(e)\\
460	&5694.7808		$_{-0.0004}^{+0.0003} $ &	0.0004	&$	0.00039	$&	89.1	$_{-0.3	}^{+0.7	}$ &	11.88	$ _{-0.17}^{+0.10}$ &	0.1402	$ _{-0.0009	}^{+0.0012	}$ &	(d)\\
567	&6038.5774		$_{-0.0006}^{+0.0009} $ &	0.0008	&$	-0.00032	$ &	89.0	$_{-0.7	}^{+0.8	}$ &	11.8	$ _{-0.7}^{+2.9	}$ &	0.124	$ _{-0.007	}^{+0.013}$ &	(c)\\
580	&	6038.5774	$_{-0.0003}^{+0.0008} $ &	0.0006	&$	-0.00089	$&	89.0	$_{-0.8	}^{+0.7	}$ &	11.2	$ _{-0.5}^{+0.3	}$ &	0.143	$ _{-0.0019	}^{+0.0028	}$ &	(g)\\
698	&6459.48784		$_{-0.00024	}^{+0.00027	} $ &	0.00026	&$	-0.00061	$&	89.0	$_{-0.5	}^{+0.6	}$ &	11.79	$ _{-0.35}^{+0.14}$ &	0.136	$ _{-0.0007}^{+	0.0018}$ &	(c)\\
783	&6732.59740		$_{-0.00020	}^{+0.00025	} $ &	0.00023	&$	-0.00107	$&	89.73	$_{-0.45}^{+0.20}$ &	11.85	$ _{-0.11}^{+0.07}$ &	0.137	$ _{-0.0007}^{+	0.0010}$ &	(c)\\
784	&6732.59740		$_{-0.0004	}^{+0.0014	} $ &	0.0009	&$-0.00133	$&	89.1$_{-0.5	}^{+0.7	}$ &12.04	$ _{-	0.30}^{+	0.17}$ &	0.130	$ _{-0.003	}^{+0.006	}$ &	(e)\\
792	&6761.51607		$_{-0.00018	}^{+0.00014	} $ &	0.00016	&$	0.00007	$&	89.77	$_{-0.59}^{+0.18}$ &	11.91	$ _{-	0.14}^{+0.06}$ &	0.140	$ _{-0.0005	}^{+0.0007	}$ &	(c)\\
793	&6732.59740		$_{-0.0006	}^{+0.0005	} $ &	0.0006	&$	-0.00126	$&	88.9	$_{-0.5	}^{+0.7	}$ &	12.02	$ _{-0.33}^{+0.18}$ &	0.137	$ _{-0.004	}^{+0.004}$ &	(g)\\
911	&7143.8694		$_{-0.0006}^{+0.0006} $ &	0.0006	&$	-0.00062	$&	88.9	$_{-0.5	}^{+0.8	}$ &	11.8	$ _{-	1.1	}^{+1.2	}$ &	0.140	$ _{-0.006	}^{+0.008}$ &	(g)\\
925	&7188.8515		$_{-0.0018	}^{+0.0027	} $ &	0.0023	&$	-0.00135	$&	89.6	$_{-1.4	}^{+0.4	}$ &	11.5	$ _{-0.8	}^{+0.6	}$ &	0.1415	$ _{-0.0019	}^{+0.0039	}$ &	(g)\\
928	&7188.8515		$_{-0.0006	}^{+0.0023	} $ &	0.0015	&$	-0.00083	$&	89.0	$_{-0.5	}^{+0.6	}$ &	11.94	$ _{-0.29}^{+0.15}$ &	0.145	$ _{-0.003}^{+0.003}$ &	(c)\\
934	&7217.7697		$_{-0.0005}^{+0.0008	} $ &	0.0007	&$	-0.00068	$&	89.0	$_{-0.9	}^{+0.7	}$ &	11.9	$ _{-0.6}^{+0.4	}$ &	0.137	$ _{-0.0018	}^{+0.0027	}$ &	(e)\\
1027	&7516.58666		$_{-0.00021	}^{+0.0011	} $ &	0.000655	&$	0.00179	$&	88.8	$_{-0.4	}^{+0.8	}$ &	11.79	$ _{-0.36}^{+0.14}$ &	0.1377	$ _{-0.0010	}^{+0.0026	}$ &	(b)\\
1045	&7574.4189 $_{-0.0003	}^{+0.0003	} $ &	0.0003	&$	-0.00103	$&	89.0	$_{-0.6	}^{+0.7	}$ &	12.08	$ _{-0.49}^{+0.13}$ &	0.1402	$ _{-0.0013	}^{+0.0020}$ &	(c)\\
1126	&7574.4189		$_{-0.0008}^{+0.0005} $ &	0.0007	&$	0.00169	$&	88.9	$_{-0.7	}^{+0.8	}$ &	12.05	$ _{-	0.49}^{+0.20}$ &	0.134	$ _{-0.0014	}^{+0.0040}$ &	(b)\\
1139&7876.4478 $_{-0.0003}^{+0.0003} $ &	0.0003	&	0.00032	& 88.9$_{-0.4}^{+0.7}$ & 11.83	$ _{-0.28}^{+0.11}$ &	0.139$ _{-0.0010}^{+0.0020}$ &	(b)\\
1140	&7879.66428		$_{-0.00028	}^{+0.00024} $ &	0.00026	&	0.00374	& 89.3$_{-1.4}^{+0.6}$ &	11.57	$ _{-0.81}^{+0.22}$ &	0.143	$ _{-0.0015	}^{+0.0060}$ &	(b)\\
1144	&7892.51414		$_{-0.00013	}^{+0.00010	} $ &	0.00012	&0.00136	& 88.9	$_{-0.22}^{+0.59}$ &	11.81	$ _{-0.15}^{+0.17}$ &	0.1375	$ _{-0.0005	}^{+0.0019	}$ &	(b)\\
1159	&7940.7071	$_{-0.0004	}^{+0.003	} $ &	0.0017	&$	-0.00156$ &	89.0	$_{-1.3	}^{+0.6	}$ &	11.83	$ _{-0.88}^{+0.18}$ &	0.139	$ _{-0.003	}^{+0.005	}$ &	(b)\\
1468	&7940.7071		$_{-0.0006	}^{+0.0009} $ &	0.0008	&-0.00110 & 88.8	$_{-1.1	}^{+0.8	}$ &	11.7	$ _{-0.8}^{+0.4	}$ &	0.1358	$ _{-0.0022	}^{+0.0040}$ &	(f)\\
1469	&8936.7555		$_{-0.0004}^{+0.0009} $ &	0.0007	&$	-0.00146$ &	89.0	$_{-1.5	}^{+0.8	}$ &	11.8	$ _{-1.0	}^{+0.4	}$ &	0.1406	$ _{-0.0017}^{+0.0048	}$ &(f)\\
1470&8936.7555		$_{-0.0004}^{+0.0009} $ &	0.0007	&$	-0.00102$ &	88.9	$_{-1.0	}^{+	0.7	}$ &	12.1	$ _{-0.7}^{+0.3	}$ &	0.1382	$ _{-0.0018	}^{+0.0035}$ &	(f)\\
1472&	8936.7555	$_{-0.0007}^{+0.0008} $ &	0.0008	&$	-0.00133	$&	89.0	$_{-1.9	}^{+0.8	}$ &	11.72	$ _{-1.25}^{+0.23}$ &	0.139	$ _{-0.003}^{+	0.006}$ &	(f)\\
1473	&8949.6075		$_{-0.0005	}^{+0.0008} $ &	0.0007	&$	-0.00169	$&	88.9	$_{-1.0	}^{+0.7	}$ &	11.8	$ _{-0.6}^{+0.4	}$ &	0.1373	$ _{-0.0018	}^{+0.0033}$ &	(f)\\
1474&8952.8210 $_{-0.0006}^{+0.0007} $ &	0.0007	&$-0.00125 $ & 	89.0	$_{-1.2	}^{+0.7	}$ &11.6 $ _{-0.87}^{+0.24}$ &	0.1383	$ _{-0.0019	}^{+0.0038}$ &	(f)\\
1672&9589.0063		$_{-0.0005}^{+0.0007} $ &	0.0006	&$	-0.00164	$ &	89.7	$_{-0.75}^{+0.22}$ &	12.29	$ _{-	0.39}^{+0.22}$ &	0.145 $ _{-0.005}^{+0.004}$ &	(g)\\
1681&9617.9237		$_{-0.0023}^{+0.0046} $ &	0.0035	&$	-0.00177	$ &	88.9	$_{-2.8}^{+0.8}$ &	11.7 $ _{-4.4}^{+5.0	}$ &	0.139	$ _{-0.043}^{+0.024}$ &	(g)\\
1689&9643.6288		$_{-0.0003}^{+0.0007} $ &	0.0005	&$	-0.00114	$ &	88.90	$_{-0.6}^{+	0.7	}$ &	11.75	$ _{-	0.42}^{+0.15}$ &	0.136	$ _{-0.0016	}^{+0.0024}$ &	(f)\\
1690&9646.8415		$_{-0.0005}^{+0.0005} $ &	0.0005	&$	-0.00150	$ &	89.0	$_{-0.6	}^{+0.7	}$ &	11.93	$ _{-	0.37}^{+0.20}$ &	0.1360	$ _{-0.0016}^{+0.0025}$ &	(f)\\
1691&9646.8415		$_{-0.0006}^{+0.0005} $ &	0.0006	& $-0.00126	$&	89.0	$_{-0.8	}^{+0.7	}$ &	11.89	$ _{-	0.45}^{+0.23}$ & 0.1345	$_{-0.0014}^{+0.0028}$ &(f)\\
1693&9646.8415		$_{-0.0005	}^{+0.0006} $ &	0.0006	& $-0.00148	$ &89.0	$_{-0.7	}^{+0.7	}$ & 12.08	$ _{-0.50}^{+	0.20}$ &	0.1348	$ _{-0.0015	}^{+0.0030	}$ &	(f)\\
1694&9659.6940	$_{-0.0004	}^{+0.0006	} $ &	0.0005	&$	-0.00124$&	89.0	$_{-0.8	}^{+0.7	}$ &	11.66	$ _{-0.54}^{+0.19}$ &	0.136 $ _{-0.0005}^{+0.0005}$ &	(f)\\
1695&9662.9065		$_{-0.0005}^{+0.0005}$ & 0.0005	& $-0.00179	$&	89.0	$_{-0.8	}^{+0.7	}$ &	11.86 $ _{-0.46}^{+	0.21}$ &	0.1370	$ _{-0.0016	}^{+0.0024	}$ &	(f)\\
1698&9672.5460	 $_{-	0.0005}^{+0.0007} $ &0.0006	& $-0.00147$ & 89.0	$_{-1.3	}^{+0.7	}$ &	11.8 $ _{-0.8}^{+0.3	}$ &	0.135 $ _{-0.0022}^{+0.0033}$ &	(f)\\
1699&9672.5460 $_{-0.0006}^{+0.0007} $ &	0.0007	&$	-0.00223$	& 89.1	$_{-1.0	}^{+0.8	}$ &	11.8	$ _{-1.0	}^{+0.3	}$ &	0.136	$ _{-0.0021}^{+0.0038}$ &	(f)\\
1702&9685.3979		$_{-0.0005	}^{+0.0006	} $ &	0.0006	& $-0.00181	$	& 89.0	$_{-0.7	}^{+0.7	}$ &	11.79	$ _{-0.42}^{+	0.21}$ &	0.1406	$ _{-0.0014	}^{+0.0028}$ &	(f)\\
1703&9688.6115		$_{-0.0008}^{+0.0006} $ &	0.0007	&$	-0.00127$ & 88.9 $_{-1.6}^{+0.3	}$ &	11.8 $ _{-1.6}^{+0.3}$ & 0.137	$ _{-0.0015}^{+0.0058}$ & (f)\\
1709&9688.6115 $_{-0.0007}^{+0.0020}$ &	0.0014	& $	-0.00412$	& 85.7	$_{-0.3	}^{+1.6	}$ & 10.4 $ _{-0.5}^{+1.5}$ &	0.158	$ _{-0.010}^{+0.006}$ &	(g)\\
\end{tabular}
\end{adjustbox}
\vspace{0.3cm}
\hspace{-0.5cm}(a)\cite{2009ApJ...706..785H},
(b)\cite{2018A&A...620A.142A},
(c)\cite{2018A&A...613A..41M},(d)\cite{2012AJ....143...95L},(e)\cite{2021RAA....21...97S},
(f)TESS,(g)our study
\end{table}

\section{Transit Timing Analysis \label{sec:TTV}}
We analyzed the TTVs of HAT-P-12b using our modeled 46 light curves. 
To investigate potential deviations from a strictly periodic transit pattern, we adopted three models: the linear model, the orbital decay model, and the Apsidal Precession model followed \cite{2017AJ....154....4P,2020AAS...23545602Y} 

\subsection{Linear model \label{subsec:linear}}
The linear model assumes a circular orbit with a constant orbital period. We derived a new ephemeris for HAT-P-12b by fitting a linear model,

\begin{equation}
    T_m^c(E) = T_0 + EP,
\end{equation}
where $T_m^c(E)$ is the calculated mid-transit time at epoch $E$, $T_0$ is the reference mid-transit time, \textit{P} is the orbital period, and \textit{E} denotes the integer epoch number. The first observed transit of HAT-P-12b is assigned $E=0$.

We used the \textit{emcee} Monte Carlo Markov Chain (MCMC) sampler \citep{2013PASP..125..306F} to fit the observed mid-transit times to the linear model. We used 50 walkers and $10^6$ steps for this analysis to sample the posterior probability distributions. The initial $10^3$ burn-in steps were discarded to ensure that the chains had to converge. We considered uniform distributions as priors for the
free parameters.

The best-fit values of the model parameters derived from the MCMC Posterior probability distribution are:
P = 3.213059  days and $T_0$ = 2454216.77325 $(BJD_{TDB})$. The resultant uniform uncertainties at one sigma are given in Table \ref{table:modelvalue}. Since the number of dimensions for the linear model is 2, the reduced chi-square $\chi^2_{red}$ with 44 degrees of freedom is 8.1. The high $\chi^2_{red}$ suggests possible deviations from strict periodicity.

\subsection{Orbital decay model}
The second model assumes a circular orbit and the period is changing at a steady rate.
\begin{equation}
    T_d^c(E) = T_{d0} + E P_d + \frac{1}{2} \frac{dP_d}{dE}E^2, 
\end{equation}
where $T_d^c(E)$ is the calculated mid-transit time, $T_{d0}$ is the mid-transit time at $E=0$, \textit{E} is the epoch number, $P_d$ is the orbital period, $\frac{dP_d}{dE}$ is the change of the orbital period in each orbit. Similar to the linear model, we used an emcee MCMC sampler to fit the orbital decay model. 

The best-fit values of the parameters from the Posterior probability distribution are P = 3.213064  days, $ T_0 $ = 2454216.77138 (BJD$_{TDB}$), and $ \frac{dP_d}{dE}$ = $-5.85_{-0.63}^{+0.64}$ x 10$^{-9}$ (days/orbit) or -0.051$\pm$0.006 seconds. The uncertainty of these parameters at one sigma is given in Table \ref{table:modelvalue}. 

The $\chi^2_{red}$ of the model with 43 degrees of freedom is 6.3 which is lower than the linear model.

\subsubsection{Stellar tidal quality}

The stellar tidal quality $ Q_\star'$ can be expressed as \citep{2018AcA....68..371M}:

\begin{equation}
 Q_\star' = -\frac{27}{2}\pi(\frac{M_p}{M_\star})(\frac{a}{R_\star})^{-5}(\frac{dP_d}{dE})^{-1}P,
 \end{equation}
where the value of $M_p$, $M_\star$ and $a/R_\star$ are taken from \cite{2009ApJ...706..785H}. The value of $ dP_d/dE $ is taken from our decay model fitting. Using these values the value of $ Q_\star $ is calculated as 28.4, a significantly lower value compared to the theoretical predictions.
The study by  \cite{2018AJ....155..165P} analyzed a sample of  188 known hot Jupiters with an orbital period $<$ 3.5 days and T$_{eff} < $ 6100 k. Their analysis estimated the value  $10^5 \leq Q_\star' \leq 10^7 $. This low $ Q_\star'$ value suggests that there must be other dynamics happening in the HAT-P-12 system that cannot be explained by the decay model. To investigate further, we choose an apsidal model for possible explanations. 

\subsection{Apsidal model}\label{apsidal_section}
The apsidal model assumes that the planetary orbit is slightly eccentric and its argument of pericenter is precessing uniformly over time \cite{2009IAUS..253..163W}. We adopted the apsidal model from \cite{1995Ap&SS.226...99G},
 
\begin{equation}
    T_{ap}^c(E) = T_{ap0} + P_sE- \frac{eP_scos(w_o + E\frac{dw}{dE})}{\pi (1-\frac{\frac{dw}{dE}}{2\pi})},
\end{equation}
where $T_{ap0}$ is the reference time, $ e $ is the eccentricity, $\omega $ is the argument of pericenter and $P_s $ is the sideral period and $ \frac{dw}{dE}$. This model has five free parameters:$T_{ap0},P_{s}, $e$,\omega $,and $\frac{dw}{dE}$.
To fit this model, we used the same MCMC sampler described for the linear model in  Section \ref{subsec:linear}. 
The best-fit values of the parameters, obtained from the Posterior probability distribution are: P= 3.213064 days, $T_0$ = 2454216.77393 (BJD$_{TDB}$),e = 0.00130, $\omega_0$ = 4.09 rad and 
$\frac{dw}{dE}$ = 0.0045 rad/epoch. 
The results of the fitted parameters with relative uncertainties are listed in the Table \ref{table:modelvalue}.
The $\chi^2_{red}$ of the model with 43 degrees of freedom is 4.2 which is better than the linear and decay model.

\subsection{Model selection}
 
The Akaike Information Criterion (AIC) and Bayesian Information Criterion (BIC) are commonly used statistical measures to determine the best model. 

\begin{equation}
    AIC = \chi^2 + 2 k,
\end{equation}
\begin{equation}
  BIC = \chi^2 + k \ln (n), 
\end{equation}
where $k$ represents the number of free parameters and $n$ denotes the number of data points. In this case, $ n = 46 $,$k = 2$ for linear model, $ k = 3 $ for decay model, $ k = 5 $ for apsidal model. 
The AIC values of the linear, decay and apsidal models are 361.19,279.71 and 185.96 respectively. Likewise, the BIC values for the linear, decay, and apsidal models are 364.85,285.19, and 195.11. The difference in  BIC between the apsidal and decay model $\Delta_{BIC}$ is 45.04. Similarly, 
the AIC favours the decay model by $\Delta_{AIC} $ is 46.87. 
To identify the most appropriate model, $\chi_{red}^2 $ and the AIC/BIC of the three models were analyzed and compared. Among these, the apsidal model indicates the lowest $\chi_{red}^2 $ value (3.1) and the lowest AIC/BIC value which 
strongly favored as the most appropriate representation of the system. 
The best-fit values of the three models,$\chi_{red}^2 $, AIC/BIC are presented in Table \ref{table:modelvalue}. The timing residuals for the orbital decay model were obtained by subtracting the mid-transit time calculated using a linear model. Similarly, the apsidal model was obtained by subtracting from the linear model mid-transit times. The resulting 0bserved-Calculated (O-C) values as a function of an epoch are illustrated in Figure \ref{fig:ocplot}.

\begin{figure*}[ht]
    \includegraphics[width=0.9\textwidth, height=10cm]{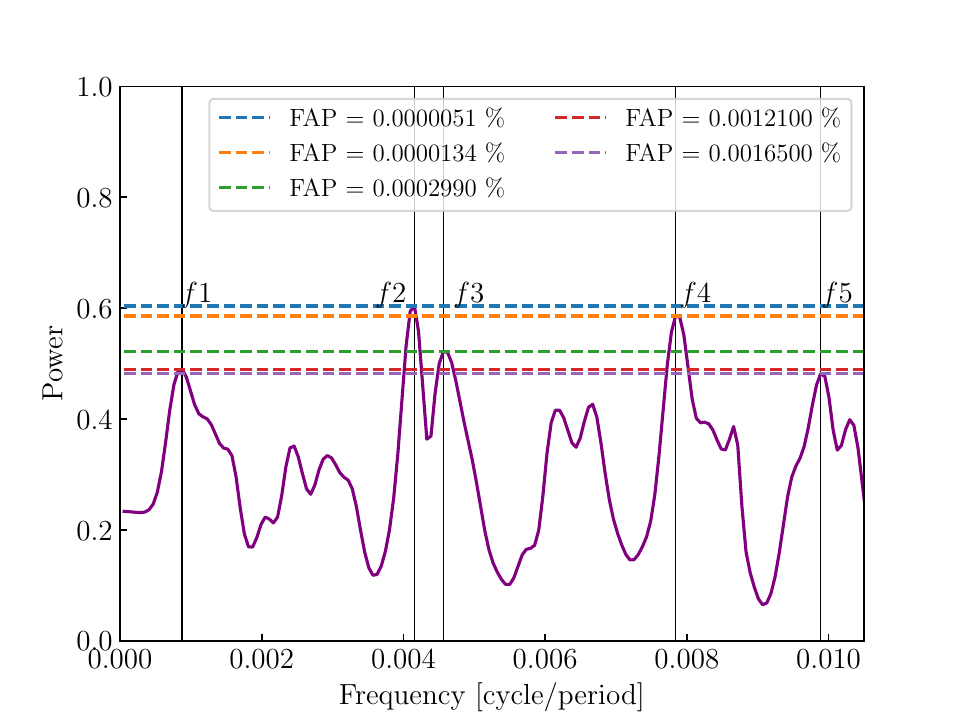}
    \caption{Generalized Lomb-Scargle periodogram of HAT-P-12b TTV data, indicating periodic signals  and their false alarm probability (FAP) levels. Peaks are annotated with their frequencies(f1 to f5)}
     \label{fig:GLS}
\end{figure*}

\begin{table}
\centering
\caption{The Uniform Priors and Best-fit Parameters for HAT-P-12b}
\vspace{0.2cm}
\label{table:modelvalue}
\begin{tabular}{cc}
\hline
\hline
Model &  Best-fit Values \\ \hline
\hline
Linear model &  \\
P[days] &  3.213059$_{-1.5\times10^{-7}}^{+1.5\times10^{-7}} $ \\
$ T_0[BJD_{TDB}-2450000] $ &  4216.77325$_{-0.00017}^{+0.00017}$\\
$\chi^2_r$ & 8.1\\
AIC/BIC & 361/364
\\
\hline
Orbital decay model &  \\
$P_d[days] $ & 3.213064$_{-6.04\times10^{-7}}^{+6.04\times10^{-7}} $ \\
$T_0[BJD_{TDB}-2450000]$ &   4216.77138$_{-0.00028}^{+0.00029}$\\
dP/dE[seconds] & -0.051$\pm$0.006 \\
$\chi^2_r$ & 6.3\\
AIC/BIC & 279/285
\\ 

\hline
Apsidal model &  \\
$P_s[days]$ &  
3.213058$_{-3.8\times10^{-7}}^{+3.8\times10^{-7}} $ \\
$T_0[BJD_{TDB}-2450000]$ &  4216.77393$_{-0.0004}^{+0.0009}$\\
e & 0.00130$_{-0.0001}^{+0.0001}$ \\
$\omega_0$[rad] &4.095$_{-0.36}^{+0.38}$ \\
dw/dE[rad/epoch] & 0.0045$_{-0.0004}^{+0.0004}$ \\
$\chi^2_r$ & 4.2\\
AIC/BIC & 185/195\\
\hline
\end{tabular} 
\end{table}

\subsection{Sinusoidal model \label{sec:frequency}}
\begin{figure*}
    \centering
    \includegraphics[width=0.9\textwidth, height=14cm]{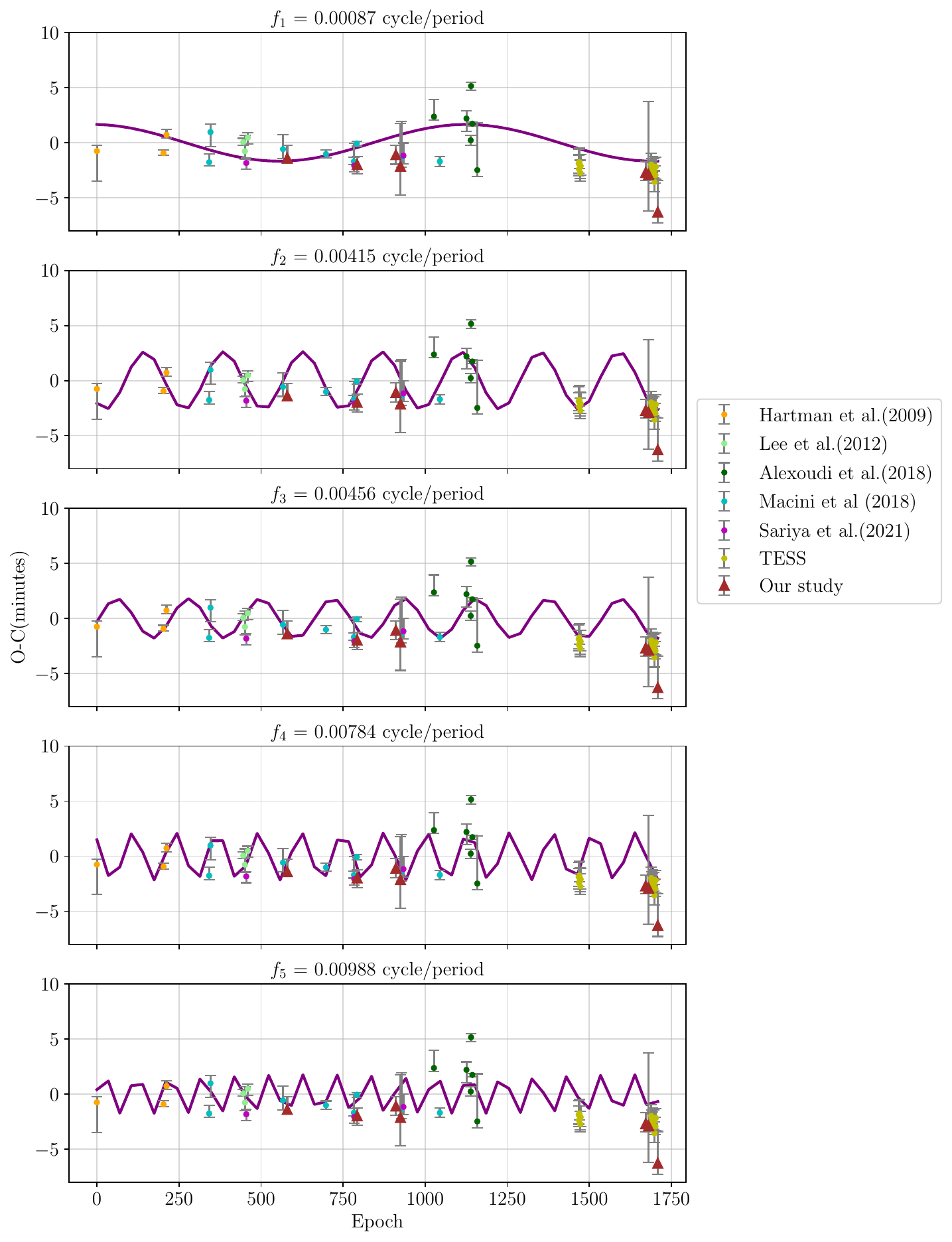}
    \caption{Sinusoidal model fitting for the most significant frequencies identified in the GLS periodogram analysis.
     \label{fig:sinusoidal model}}
\end{figure*}

To explore the possibility of additional planets affecting the Transit times of HAT-P-12b, we analyzed the observed TTVs using the Generalized Lomb-Scargle  (GLS) Periodogram
\citep{1976Ap&SS..39..447L,1982ApJ...263..835S}. This method is particularly well-suited for analyzing unevenly spaced time series data, which is common in transit observations. It offers the advantage of incorporating uncertainties, thereby improving the robustness of the frequency analysis \citep{2009A&A...496..577Z}.
We used the GLSP code provided by the \textit{PyAstronomy\footnote{\url{https://github.com/sczesla/PyAstronomy}}}
package. The default normalization Zechmeister \& Kurster (ZK) was used for normalization. 
The resulting periodogram shown in Figure \ref{fig:GLS} exhibited multiple peaks of nearly the same height.
We used \textit{$find\_peaks$} from the scipy \footnote{\url{https://docs.scipy.org/doc/scipy/reference/generated/scipy.signal.find_peaks.html}} package to find the top five prominent peaks and labeled as
$f1$ to $f5$ and consider their false alarm probability (FAP) to check their significance. The frequency of each peak and their FAP values are summarized in Table \ref{table:frequency}. 
To evaluate these peaks further, we fit each one with a sinusoidal model given by \cite{2019A&A...628A.116V},

\begin{equation}
    TTVs(E) = A_{TTVs}\sin(2\pi fE-\phi),
\end{equation}
where $A_{TTVs}$ is the amplitude of TTV in minutes, 
$E$ is the epoch, and 
$\phi$ is the phase, $f$ is the frequency. For each frequency value obtained from the GLS periodogram, the sinusoidal fitting was performed and we estimated the values for $\phi $ and $A_{TTVs}$. We calculated the  $\chi^2_R$ and BIC for these fits to choose the best frequency. The peak at f2 = 0.00415 cycles/day had the lowest values of $\chi^2_r$ =3.2, BIC =149.3 and lowest FAP = $0.5\times 10^{-5}$. The amplitude of the best frequency is A$_{TTV}$ is 2.6 minutes.
The FAPs, along with the amplitude, phase, $\chi^2_r$, and BIC values for each frequency f1-f5 are presented in Table \ref{table:frequency}. Our timing residuals(O-C) clearly show the sinusoidal pattern, which indicates the presence of periodic signal are shown in Figure \ref{fig:sinusoidal model}. This strongly suggests the influence of an additional planet in the system. The sinusoidal fitting allowed us to estimate the parameters of the additional planet. In our next section, we aim to find the mass and period of the candidate planet.

\begin{table}
\centering
\caption{Frequency values of sinusoidal model}
\label{table:frequency}
\begin{tabular}{ccccccc}
\hline
\hline
No.of &frequency& FAP& $A_{TTV}$& Phase & $ \chi_r^2 $ & BIC \\
frequency& (cycle/period)& & (minutes) & & & \\
\hline
\hline
f1 &0.00087 & 0.00121 & 1.64 $_{-0.12}^{+0.12}$ & 1.49$_{-0.09}^{+0.09}$ & 4.21 & 193.0 \\
f2 &0.00415 & 0.000005 &2.62 $_{-0.19}^{+0.19}$& 2.11$_{-0.05}^{+0.04}$ & 3.22 & 149.3 \\
f3& 0.00456&  0.000299 & 1.80$_{-0.14}^{+0.13}$& 0.162$_{-0.08}^{+0.08}$ & 3.90 & 179.3\\
f4& 0.00784& 0.0000134 & 2.15$_{-0.18}^{+0.17}$& 0.76$_{-0.06}^{+0.06}$ & 3.36 & 155.5\\
f5& 0.00988& 0.00165 & 1.75$_{-0.15}^{+0.14}$& 0.13$_{-0.10}^{+0.10}$ & 4.26 & 195.4\\
\hline
\end{tabular}
\end{table}

\subsection{Mass of an additional planet}

To estimate the mass of the perturbing planet, we employed the analytical model of \cite{2012ApJ...761..122L}
which relates the TTV amplitude (V) to the orbital period, $\mu$ is the mass of outer planet, and j is the resonance. 

\begin{equation}
      V = P\frac{\mu}{ \pi j^{2/3}(j-1)^{1/3}\Delta}(-f-\frac{3}{2}\frac{Z_{free}^*}{\Delta}), 
\end{equation}
where $\Delta = \frac{P'}{P} \frac{j}{j-1}-1 $, represents the normalized distance from resonance, $f$ is the Laplace coefficients, $Z^*_{free}$ is the free eccentricity, respectively. We assume that the perturber planet is
in a 2:1 MMR j:j-1 with HAT-P-12b and both planets have circular orbits, 
i.e. $Z^*_{free}$ = 0.

Using the TTV amplitude found from our sinusoidal model ($V$=2.6 minutes) and the known orbital period ($P$) of HAT-P-12b  we estimated the perturbing planet's mass to be approximately 0.02 $M_J$ and its orbital period to be approximately 6.24 days.

\subsection{Applegate mechanism}

We examined the Applegate mechanism to determine whether the observed TTVs in the 
HAT-P-12 system could result from stellar activity \citep{1992ApJ...385..621A}.
The stellar activities of 59 systems including HAT-P-12b were analyzed by \cite{2010MNRAS.405.2037W} and found that the amplitude of the Applegate effect over 11, 22, and 50-year timescales. They found that for HAT-P-12b, $\delta t_{50}$ = 0.4 s. 
However, our observed TTV amplitude found using the sinusoidal model with the frequency f2
(see section \ref{sec:frequency}) is 156 seconds, significantly larger than the amplitude of the Applegate effect.

The variations due to the Applegate mechanism are at best quasiperiodic, while the TTVs caused by an additional planet in a mean-motion resonance orbit generally have strictly periodic and larger transit-time variations. 
It is obvious that our observed TTVs exhibit sinusoidal variations as shown in Figure \ref{fig:sinusoidal model}.
Therefore Applegate mechanism may not be the possible cause of TTVs in the HAT-P-12 system.

\section{The Summary and Concluding Remark\label{sec:RD}}

This study investigates the Transit Timing Variations (TTVs) of the sub-Saturn mass planet HAT-P-12b around a metal-poor K4 dwarf star. By examining 46 transit light curves across a 
14-year baseline, we analyze possible explanations for the variations in transit timings. We used four models to study TTVs: a linear, decay model, apsidal precession, and sinusoidal model.
First, the transit timings were modeled using a linear model. This model assumes a constant orbital period and provides a baseline for detecting deviations. The $\chi^2_r$ = 8.1 showed that the timing residuals exhibited non-random patterns. This finding strongly suggested that the orbit of HAT-P-12b has been affected by additional forces or interactions. 
We examined the possible effects of orbital decay, i.e. a steady decrease of the orbital period induced by tidal dissipation between the planet and its host star. The orbital decay model produced a lower $\chi^2_r$ = 6.3, suggesting a better fit than the linear model.
However, the computed stellar tidal quality 
($Q_\star \sim 28.4$) was substantially below theoretical estimates
$10^5 \leq Q_\star \leq 10^7$ for stars of similar properties.  Therefore orbital decay is not the dominant mechanism in the HAT-P-12 system. The apsidal precession model, which incorporates a slight orbital eccentricity and a precession in the argument of periastron, provided the significantly lower $\chi^2_r$ = 4.2 to the timing residuals than the linear and the decay model.
The results show, a small eccentricity e = 0.00130 and the precession $\frac{dw}{dE}$ = 0.0045 radians per epoch. 
To explore the possibility of a second planet in the system, we used the GLS periodogram to do a frequency analysis. 
Periodic signals that are not uniformly spaced in time-series data, such as transit times, can be found using this  method. The GLS periodogram showed many prominent peaks, the most prominent of which corresponded to a frequency of f2 = 0.00415 cycles per day. With a false alarm probability (FAP) of 
0.5$\times$10$^{-5}$, this peak showed a highly significant periodic signal.
To investigate further, we used a sinusoidal model to fit the timing residuals using the frequency values obtained from the GLS periodogram. The sinusoidal fit shows a lower 
$\chi^2_r$ = 3.2 indicating the best model for the timing residuals. The amplitude of the best frequency $A_{TTV}$ = 2.6 minutes. The periodic nature of the f2 signal strongly suggests the additional planetary companion in the HAT-P-12 system. 
Utilizing the amplitude and period of the sinusoidal signal, we estimated the mass of the perturbing planet to be approximately 0.02 $M_J$, assuming a 2:1 Mean Motion Resonance (MMR). To verify that the timing residuals are not caused by stellar activity, we investigated the Applegate effect. However, this results in an amplitude of only 0.4 s, significantly less than the measured TTV amplitude of 156 seconds. Furthermore, the quasiperiodic changes usually associated with stellar activity contrast the truly periodic behavior of the observed signal. We therefore conclude that the observed TTVs are not likely to be caused by stellar activity.
The perturbing planet's orbital period and estimated mass suggest a low-mass companion that could be located in a 2:1 mean-motion resonance with HAT-P-12b. We suggest that long-term observation and precise timing are required to discover the complex gravitational forces influencing planetary orbits.
The work on HAT-P-12 could offer new perspectives on the dynamics of low-mass companions in sub-Saturn systems.

\vspace{3mm}

\section*{Acknowledgements}
We are grateful to the anonymous referee for good suggestions.
This project is supported in part 
by the National Science and Technology Council, Taiwan, under
Ing-Guey Jiang's 
Grant NSTC 111-2112-M-007-035, NSTC 113-2112-M-007-030,
and also Li-Chin Yeh's
Grant NSTC 113-2115-M-007-008. 
P.T. expresses his sincere thanks to IUCAA, Pune, for providing support through the IUCAA Associateship Programme.  V.K.M. acknowledge Govt. Niranjan Kesharwani College, Kota, Bilaspur (C.G.) for providing research facilities in the college.

\bibliographystyle{elsarticle-harv} 
\bibliography{hatp12b}


\end{document}